# Is Complex Probability Theory Consistent with Bell's Theorem?

Saul Youssef

Supercomputer Computations Research Institute
Florida State University
Tallahassee, Florida 32306–4052
youssef@scri.fsu.edu

## Abstract

Bayesian complex probability theory is shown to be consistent with Bell's theorem and with other recent limitations on local realistic theories which agree with the predictions of quantum mechanics.

Quantum mechanics is often introduced with a discussion of the two slit experiment where the observed interference pattern forces us to conclude that a single particle goes through both slits at once and is "both a particle and a wave." This most basic argument has a loophole, however. The conclusion rests on probability theory, and, in particular, on the fact that probabilities are non-negative so that when the second slit is opened,

$$P(x) = P(x \text{ via slit 1}) + P(x \text{ via slit 2}) \geq P(x \text{ via slit 1}) \tag{1}$$

where $P(x)$ is the probability for the particle to arrive at position $x$ on the screen. Perhaps, then, it is possible that the particle does go through either one slit or the other after all and the interference effects can be explained by modifying probability theory itself. This possibility is put forward in references 1 and 2 where a Bayesian complex probability is formulated and shown to have a frequency interpretation, to imply a superposition principle, to include wavefunctions which are expansions in eigenfunction of Hermitian operators, to have a path integral representation, to describe both pure and mixed systems and to imply the Schrödinger equation for a scalar particle. Since this formulation is both realistic and local (as defined below), coexistence with Bell's theorem and other limitations on local realistic theories is an issue. Although coexistence with Bell's result has already been discussed[1,2], here we present the argument in more detail and also consider other more recent limitations on local realism.

One can view the net effect of quantum mechanics to be an assignment of complex numbers to pairs of propositions. The complex probability that proposition $b$ is true given proposition $a$ is denoted by "$(a \rightarrow b)$." Propositions which specify a time will often be denoted by a subscript, so, for example, $a_t$ means "$a$ is true at time $t$." Such an arrow is a *quantum theory* if[3]

I. The arrow is a *complex probability*:

$$(a \rightarrow b \wedge c) = (a \rightarrow b)(a \wedge b \rightarrow c) \tag{I.a}$$
$$(a \rightarrow b) + (a \rightarrow \neg b) = 1 \tag{I.b}$$
$$(a \rightarrow \neg a) = 0 \tag{I.c}$$

for all propositions $a$, $b$ and $c$ and if

II. There exists a set $U$ (the *state space*) and a real *time* $t$ such that:

$$x_t \wedge y_t = false \text{ if } x \neq y \tag{II.a}$$
$$(a_t \rightarrow b_{t''}) = (a_t \rightarrow U_{t'} \wedge b_{t''}) \tag{II.b}$$
$$(a_t \wedge x_{t'} \rightarrow b_{t''}) = (x_{t'} \rightarrow b_{t''}) \tag{II.c}$$

for all $x, y \in U$, for all propositions $a, b$ and for all times $t \leq t' \leq t''$ with $U_t \equiv \vee_{x \in U} x_t$.

It is convenient to call a proposition $q$ *normal* if $\int_{x \in U} |q_t \rightarrow x_{t'}|^2$ exists and is greater than zero for all $t' \geq t$ and to let a time subscript on a set of propositions denote the *or* of all its elements: $W_t = \vee_{w \in W} w_t$. Following probability theory, propositions $a$ and $b$ are said to be *independent* if $(q \wedge a \rightarrow b) = (q \rightarrow b)$ for all propositions $q$. Also, please note that in spite of the fact that we call $U$ the "state space," $U$ is an ordinary measure space and is not a Hilbert space.

As discussed in references 1 and 2, the Bayesian consistency conditions[4] for complex probabilities lead uniquely to axioms I.a–I.c which have exactly the same form as ordinary Bayesian probability theory except that the probabilities are complex, while (II.a–II.c) allow a frequency meaning to be derived for complex probabilities. Intuitively, II is just an expression of what is meant by "the system has a state": II.a guarantees that the system cannot be in two states at once, II.b guarantees that a system is in some state at each intermediate time and II.c means that if a system is known to be in some particular point in state space, all previous knowledge of the system is irrelevant. For example, in the two slit experiment, if a particle is detected at the screen, this implies $U_t$ for any previous time, or, more generally, that the particle reached the screen by some definite but unknown path and came through exactly one of the two slits. Thus, these axioms have some of the flavor of a hidden variable theory where the "hidden" variable is the particles's position but with no assumption that knowledge of $x_t \in U_t$ determines the system's future. The point of this paper is to show that the proposed modification of probability theory allows such theories to escape Bell's theorem and other Bell–like results.

As with ordinary Bayesian probabilities, complex probabilities are not defined in terms of frequencies; a frequency meaning must be derived after the fact. Similarly, one can show that in a quantum theory $U$, if normal $e$ is known at time $t$, then

$$Prob(e_t, a_{t'}) = \frac{\int_{x \in U} |e_t \to a_{t'} \wedge x_{t'}|^2}{\int_{x \in U} |e_t \to x_{t'}|^2} \tag{2}$$

gives the ordinary probability that $a$ will be true at time $t'$. Also, just as in standard probability theory, locality enters via assumptions of statistical independence[5]. For example, if two unrelated experiments $e_1$ and $e_2$ have possible results $r_1$ and $r_2$ respectively, then the assumptions that $\{r_1, r_2\}$, $\{e_1, r_2\}$ and $\{e_2, r_1\}$ are independent allow the expected conclusion $(e_1 \wedge e_2 \to r_1 \wedge r_2) = (e_1 \to r_1)(e_2 \to r_2)$. Whether a quantum theory is local or not depends purely upon such added assumptions. However, the usual assumption that propositions about events with space–like separation are independent leads to the same predictions as quantum mechanics[1,2]. Thus, we have a situation where quantum theories are: local in the sense just explained, realistic in the sense of axiom II and agree with the predictions of quantum mechanics. This would seem to be in conflict with Bell's theorem[6] and other limitations on local realistic theories[7-10]. Although space does not permit the discussion of all such results here, we show that no conflict exists for a representative subset of Bell–like results.

Before considering Bell's original result it is useful to carry (2) slightly further and say that a set of propositions $W$ *supports probabilities with initial knowledge* $e_t$ if, for all $t' \geq t$, for all $A, B \subset W$,

$$A_{t'} \wedge A^c_{t'} = false \text{ and } U_{t'} \text{ implies } W_{t'} \tag{3}$$

$$Prob(e_t, A_{t'} \wedge B_{t'}) = Prob(e_t, A_{t'}) \, Prob(e_t \wedge A_{t'}, B_{t'}) \tag{4}$$

$$Prob(e_t, A_{t'}) + Prob(e_t, A_{t'}^c) = 1 \tag{5}$$

where the complement $A^c = W - A$. The conditions under which $W$ supports probabilities can be easily found.

**Theorem 1.** *If $U$ is a quantum theory, $e_t$ is a normal proposition and $W$ is a set of propositions satisfying equation (3) above, then $W$ supports probabilities with initial knowledge $e_t$ if and only if, for all times $t' \geq t$ and for all subsets $A$ of $W$, $\int_{x \in U} (e_t \to A_{t'} \wedge x_{t'})^* (e_t \to A_{t'}^c \wedge x_{t'}) + (e_t \to A_{t'} \wedge x_{t'})(e_t \to A_{t'}^c \wedge x_{t'})^* = 0$. (Proof in reference 2).*

For any set of propositions satisfying Theorem 1, including the state space itself, "$Prob$" in equations (4)–(5) above has the frequency meaning of ordinary probabilities.

In Bell's analysis[6], two spin $\frac{1}{2}$ particles in a singlet state are emitted towards two distant Stern–Gerlach magnets. Let $e_t$ define the known orientations of the two magnets and the description of the initial singlet state and let $M_{t''}$ be a description of one of the possible results of the final measurements. Let $t$ be the time when the singlet state is created, $t''$ be the time of the final measurement and let $t < t' < t''$. Bell's theorem is an argument in probability theory beginning with an expansion in "hidden variable" $\lambda$ in state space $U$: $P(e_t, M_{t''}) = P(e_t, U_{t'} \wedge M_{t''})$ and so

$$P(e_t, M_{t''}) = \int_{\lambda \in U} P(e_t, \lambda_{t'} \wedge M_{t''}) = \int_{\lambda \in U} P(e_t, \lambda_{t'}) P(e_t \wedge \lambda_{t'}, M_{t''}). \tag{6}$$

However, since ordinary probability theory has been abandoned, equation (6) must be justified within complex probability theory. But if the definition of $Prob$ is extended to mixed times where for normal $e_t$ with $t \leq t' \leq t''$ we let

$$Prob(e_t, A_{t'} \wedge B_{t''}) = \frac{\int_{x \in U} |e_t \to A_{t'} \wedge B_{t''} \wedge x_{t''}|^2}{\int_{x \in U} |e_t \to x_{t''}|^2} \tag{7}$$

then

$$Prob(e_t, A_{t'} \wedge B_{t''}) = \frac{\int_U |e_t \to A_{t'} \wedge x_{t''}|^2}{\int_U |e_t \to x_{t''}|^2} Prob(e_t \wedge A_{t'}, B_{t''}) \tag{8}$$

does not generally satisfy (4) unless $t' = t''$. In particular, $\int_{x \in U} |e_t \to A_{t'} \wedge x_{t''}|^2 = \int_{x \in U} |\int_{y \in U} (e_t \to A_{t'} \wedge y_{t'})(y_{t'} \to x_{t''})|^2$ and, if one could take the square inside the integral, (4) could be satisfied. Since this fails because of the "interference terms" involved in exchanging the square with the integral, an appropriate classical limit might restore (3)–(5) for propositions with mixed times. However, within complex probability, we are unable to justify this crucial step in Bell's analysis. Thus, although Bell's theorem is usually interpreted as ruling out local realistic theories, in a more general context Bell's result actually forces a choice between local realism and standard probability theory.

In order to see the escape from Bell's result more directly, consider one of the clearest presentations of it, due to Mermin[7]. In this variation on Bell's result, two electrons in a singlet state enter two distant Stern–Gerlach magnets as usual. Each magnet can be set to measure spin along one of three axes at 0, 120 or 240 degrees perpendicular to the flight path of the electrons. Supposing that both electrons have an internal state which determines the result of each of the three measurements, let $a, b, c \in \{-1, 1\}$ be the predetermined results of the three possible measurements for one of the magnets and let $a', b', c' \in \{-1, 1\}$ be the corresponding results for the remaining magnet. Mermin then shows that there is no sequence of 6–tuples $(a, b, c, a', b', c')$ which reproduces the experimentally verified predictions of quantum mechanics and therefore such an assignment of predetermined values is impossible. Notice, however, that such a sequence exists if and only if there is a standard probability theory with sample space $\{(a, b, c, a', b', c') : a, b, \ldots \in \{-1, 1\}\}$ which agrees with the predictions of quantum mechanics[11]. For a set of propositions which do not satisfy the orthogonality condition of theorem 1, there is no corresponding standard probability theory and so no corresponding sequence exists. Thus, we have the following remarkable situation:

(a) Any of the possible measurement results has a well defined probability with the usual frequency interpretation;
(b) However, there is no sequence: $(a, b, c, a', b', c'), \ldots$ which represents these predictions;
(c) Nevertheless, any set of propositions satisfying theorem 1, including the state space, do support conventional probabilities and therefore can be represented by a sequence of values.

We have implicitly assumed that the propositions "spin up 0," "spin up 120" etc. do not satisfy the orthogonality condition of theorem 1, but this is quite plausible. Thus, Mermin's analysis is correct and it is not possible to assign values to $(a, b, c, a', b', c')$. However, it remains possible to assign values to other sets of propositions satisfying theorem 1 including the state space.

Another class of limitations on local realism is inspired by recent results showing that local hidden variable theories are inconsistent with certain "perfect correlations" predicted in standard quantum mechanics[8]. In the context of this analysis, our quantum theory is a hidden variable theory where the unknown actual paths of the particles through state space constitute the hidden variables. Consider, for example, the triple armed interferometer proposed in reference 8 with phase shifts $\phi_1$, $\phi_2$ and $\phi_3$ for the three arms and where each arm ends in a beam splitter directing the particle either into primed or unprimed detectors (Figure 1). Following reference 8, for hidden variable $\lambda$, let $A_\lambda(\phi_1)$, $B_\lambda(\phi_2)$ and $C_\lambda(\phi_3)$ be $+1$ if the respective unprimed detector responds and $-1$ if the respective primed detector responds. One can then show that

$$\text{if } \phi_1 + \phi_2 + \phi_3 = \pi/2 \text{ then } A_\lambda(\phi_1)B_\lambda(\phi_2)C_\lambda(\phi_3) = +1 \text{ and} \qquad (9)$$

$$\text{if } \phi_1 + \phi_2 + \phi_3 = 3\pi/2 \text{ then } A_\lambda(\phi_1)B_\lambda(\phi_2)C_\lambda(\phi_3) = -1 \qquad (10)$$

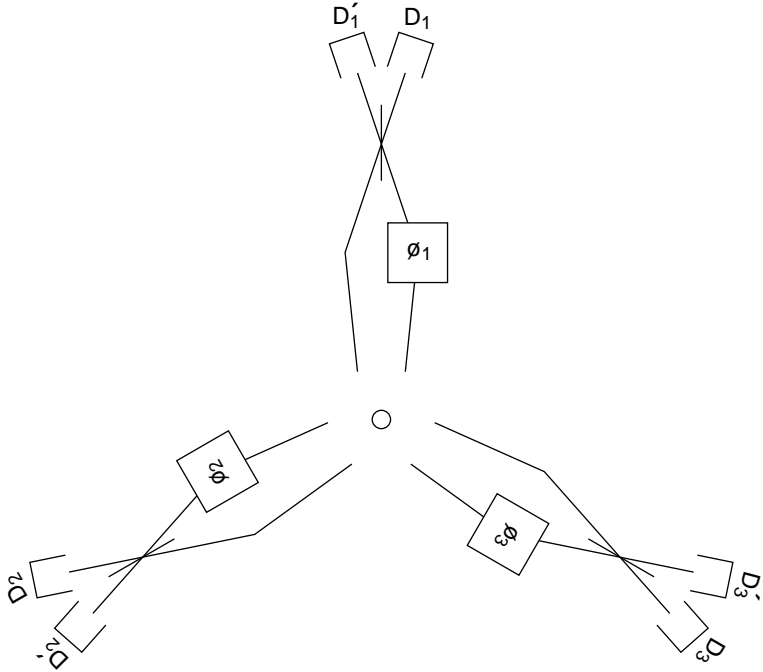

*Figure 1. The three armed interferometer proposed in reference 8 consists of a central source which emits three particles in the plane of the paper with zero total momentum. Each particle is brought to a beam splitter and is detected in either a primed or unprimed detector ($D'_j$ and $D_j$ respectively). For each arm of the interferometer, one of its paths induces an adjustable phase shift $\phi_1$, $\phi_2$ or $\phi_3$ as indicated.*

and, by choosing particular phase arrangements, that $A_\lambda(\pi/2)B_\lambda(0)C_\lambda(0)$, $A_\lambda(0)B_\lambda(\pi/2)C_\lambda(0)$, and $A_\lambda(0)B_\lambda(0)C_\lambda(\pi/2)$ are all equal to $+1$. But since the product of these factors is $A_\lambda(\pi/2)B_\lambda(\pi/2)C_\lambda(\pi/2) = +1$, a contradiction with (10) is reached and such hidden variable theories are ruled out. However, this analysis assumes that a single $\lambda$ can occur with any of the chosen phase settings. This is not so in our case where the "hidden variables" are the actual paths of the three particles through the apparatus. In particular, one can verify (Ref. 8, Appendix G) that no possible experimental final state occurs both for a phase setting satisfying $\phi_1 + \phi_2 + \phi_3 = \pi/2$ and for a phase setting satisfying $\phi_1 + \phi_2 + \phi_3 = 3\pi/2$. Thus no $\lambda$ value satisfies both (9) and (10) and a contradiction is avoided.

Another type of non–locality result has been proposed[9] using two overlapping interferometers with paths $P_j^+$, $P_j^-$ (before point $P$) and $Q_j^+$, $Q_j^-$ (after point $P$) with detectors $D_j^+$ and $D_j^-$ for $j = 1, 2$ (figure 2). Electrons and positrons encounter two beam splitters as they pass from the source to the detectors. The second beam splitter in each of the electron and positron arms is optionally removable. The phases of the beam splitters are adjusted so that if electrons or positrons enter the device separately, there are zero counts in detectors $D_1^\pm$.

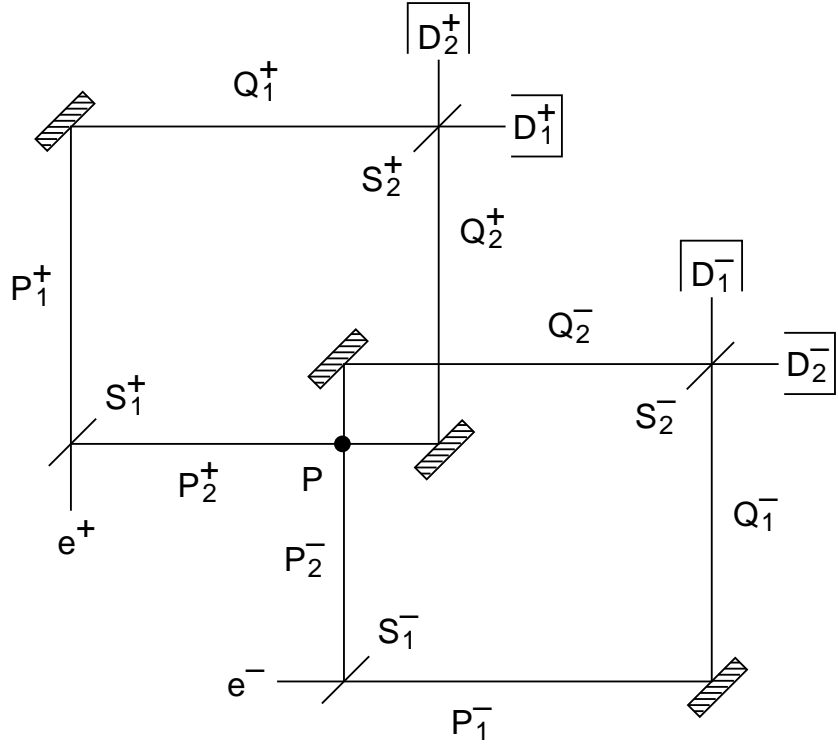

*Figure 2. Two overlapping interferometers as described in reference 11. Electrons and positrons first encounter a beam splitter ($S_1^\pm$), then one of two mirrors, a second beam splitter ($S_2^\pm$) and are then detected in one of $D_j^\pm$. The electron and positron are assumed to annihilate with probability one if they meet at point P.*

Following reference 9, assume that if a positron and electron meet at point $P$, they annihilate with probability one. This situation can be described with a quantum theory with state space $U = (\mathbb{R}^3 \times \mathbb{R}^3) \cup \Omega$ where $\Omega$ is the state space of the electron–positron decay products. Let $E$, $E^+$, $E^-$ and $E^\pm$ be the initial descriptions of experiments with no second beam splitters, just the positron second beam splitter, just the electron second beam splitter and with both second beam splitters respectively. Then, suppressing time subscripts and using axiom II.b twice gives

$$(\mathcal{E} \to D_p^+ \wedge D_q^-) =$$

$$(\mathcal{E} \to \{(P_1^+ \vee P_2^+) \wedge (P_1^- \vee P_2^-)\} \wedge \{\Omega \vee ((Q_1^+ \vee Q_2^+) \wedge (Q_1^- \vee Q_2^-))\} \wedge D_p^+ \wedge D_q^-)) \quad (11)$$

for $\mathcal{E}$ equal to any of $E$, $E^+$, $E^-$ or $E^\pm$. Since $\Omega \wedge D_j^\pm = false$,

$$(\mathcal{E} \to D_p^+ \wedge D_q^-) = \sum_{j,k,l,m=1}^{2} (\mathcal{E} \to P_j^+ \wedge P_k^- \wedge Q_l^+ \wedge Q_m^- \wedge D_p^+ \wedge D_q^-) \quad (12)$$

and using II.c,

$$(\mathcal{E} \to D_p^+ \wedge D_q^-) = \sum_{j,k,l,m=1}^{2} (\mathcal{E} \to P_j^+ \wedge P_k^-)(P_j^+ \wedge P_k^- \to Q_l^+ \wedge Q_m^-)(Q_l^+ \wedge Q_m^- \to D_p^+ \wedge D_q^-). \quad (13)$$

Assuming that $P_j^+$ and $P_k^-$ are independent, $(\mathcal{E} \to P_j^+ \wedge P_k^-) = (\mathcal{E} \to P_j^+)(\mathcal{E} \to P_k^-)$. Similarly, assuming that $\{D_p^+, D_q^-\}$, $\{Q_l^+, D_q^-\}$ and $\{Q_m^-, D_p^+\}$ are independent,

$$(Q_l^+ \wedge Q_m^- \to D_p^+ \wedge D_q^-) = (Q_l^+ \to D_p^+)(Q_m^- \to D_q^-). \quad (14)$$

Then with[3] $(P_j^+ \wedge P_k^- \to Q_l^+ \wedge Q_m^-) = \delta_{jl}\delta_{km}\lceil \neg(j = k = 2)\rceil$,

$$(\mathcal{E} \to D_p^+ \wedge D_q^-) = \sum_{j,k=1}^{2} (\mathcal{E} \to P_j^+)(\mathcal{E} \to P_k^-)(Q_j^+ \to D_p^+)(Q_k^- \to D_q^-)\lceil \neg(j = k = 2)\rceil. \quad (15)$$

Without loss of generality, we can choose $(\mathcal{E} \to P_j^\sigma) = e^{i\lceil j=2\rceil \pi/2}$, $(Q_j^\sigma \to D_p^\sigma) = \lceil j = p\rceil$ (if beam splitter $\sigma$ is removed) and $(Q_j^\sigma \to D_p^\sigma) = e^{i\lceil j \neq p\rceil \pi/2}$ (if beam splitter $\sigma$ is in place) which is consistent with $D_1^\pm$ giving zero counts if electrons or positrons enter the device separately. Substituting these conditions into equation (15) reproduces equations 9–13 of reference 9. Thus, the predictions of standard quantum mechanics also follow in our quantum theory.

Following reference 9, introduce hidden variable $\lambda$ which, in this case, is the actual paths of the two particles through the experiment and which determines $D_j^\pm$ independent of which beam splitters are in place. Then

$$D_2^+(\lambda) \wedge D_2^-(\lambda) = false \text{ for experiment } E, \quad (16)$$

$$\text{if } D_1^+(\lambda) \text{ then } D_2^-(\lambda) \text{ for experiment } E^+ \text{ and} \quad (17)$$

$$\text{if } D_1^-(\lambda) \text{ then } D_2^+(\lambda) \text{ for experiment } E^- \quad (18)$$

and consider an instance of experiment $E^\pm$ with $D_1^+(\lambda) \wedge D_1^-(\lambda) = true$. It is argued that this, together with (17) and (18) implies $D_2^+(\lambda) \wedge D_2^-(\lambda) = true$, contradicting (16). However, since $(E^+ \to D_1^+ \wedge D_1^-) = 0$, there is no $\lambda$ which is both a possible path in experiment $E^+$ and which also satisfies $D_1^+(\lambda) \wedge D_1^-(\lambda) = true$. Thus, a contradiction is once again avoided.

I acknowledge Angel Amilibia, Mary-Anne Cummings, Chris Georgiopoulos, Asher Klatchko, Arthur Kosowsky, Daryl McCullough, Tim S. Roberts and Richard Silver for helpful comments.

# References

[1] S. Youssef, Mod.Phys.Lett. 6 (1991) 225.

[2] S. Youssef, Mod.Phys.Lett. 9 (1994) 2571.

[3] We use this standard notation for boolean operations: $\wedge$ for *and*, $\vee$ for *or* and $\neg$ for *not*. $P(a,b)$ in our notation is often written $P(b|a)$. If $A$ is a proposition, $\lceil A \rceil$ denotes the function which is 1 if $A$ is *true* and 0 if $A$ is *false*.

[4] R.T.Cox, Am.J.Phys. 14 (1946) 1.

[5] It is trivial to show that for propositions which support probabilities with some initial knowledge, independence in the complex sense implies statistical independence in the usual sense.

[6] J.S.Bell, Physics, 1 (1964) 195; J.S.Bell, Rev.Mod.Phys. 38 (1966) 447. See J.Bub, Found.Phys. 3 (1972) 29 for comments and N.D.Mermin, Rev.Mod.Phys. 65 (1993) 803 for a recent review.

[7] N.D.Mermin, Am.J.Phys. 49 (1981) 10.

[8] D.M.Greenberger, M.A.Horne, A.Shimony and A.Zeilinger, Am.J.Phys. 58 (1990) 1131.

[9] L.Hardy, Phys.Rev.Lett. 20 (1992) 2981.

[10] E.P.Wigner, Am.J.Phys. 38 (1969) 1005; J.F.Clauser and M.A.Horne, Phys.Rev. D 10 (1974) 526; P.H.Eberhard, Nuovo Cimento 38 B (1977) 75; J.D.Franson, Phys.Rev.Lett. 62 (1989) 2205; N.D.Mermin, Phys.Rev.Lett. 65 (1990) 3373; A.Peres, Phys.Lett. A 151 (1990) 107; L.Hardy, Phys.Lett. A 161 (1991) 21; I.Pitowsky, Phys.Lett. A 156 (1991) 137; A.C.Elitzur, S.Popescu and D.Rohrlich, Phys.Lett. A 162 (1992) 25; L.Hardy, Phys.Lett. A 167 (1992) 17; L.Hardy and E.J.Squires, Phys.Lett. A 168 (1992) 169; A.Mann, K.Nakamura and M.Revzen, J.Phys. A25 (1992) L851; C.Pagonis and R.Clifton, Phys.Lett. A 168 (1992) 100; M.Vinduska, Found.Phys. 22 (1992) 343; A.Elby and M.R.Jones, Phys.Lett. A 171 (1992) 11; M.Ardehali, Phys.Lett. A 181 (1993) 187; M.Ardehali, Phys.Rev. A 47 (1993) 1633; H.J.Bernstein, D.M.Greenberger, M.A.Horne and A.Zeilinger, Phys.Rev. A 47 (1993) 78; P.Busch, P.Kienzler, P.Lahti and P.Mittelstaedt, Phys.Rev. A 47 (1993) 4627; D.N.Klyshko, Phys.Lett. A 172 (1993) 399; H.P.Stapp, Phys.Rev. A 47 (1993) 847; B.Yurke and D.Stoler, Phys.Rev. A 47 (1993) 1704; M.Czachor, Phys.Rev. A 49 (1994) 2231; L.Hardy, Phys.Rev.Lett. 73 (1994) 2279.

[11] For if such a probability exists, you can generate a sequence by Monte Carlo sampling; conversely, an infinitely long sequence determines all such probabilities. Also see A.Fine, Phys.Rev.Lett. 48 (1982) 291.